\documentclass{article}

\usepackage{spconf}
\usepackage[short]{longShort} 
\usepackage{graphicx}
\graphicspath{{imgs/}}
\usepackage{tikz}
\usetikzlibrary{shapes.misc}
\usetikzlibrary{shapes.geometric}
\usepackage{mathtools}
\usepackage{commath}
\usepackage{amssymb}
\usepackage{iftex}
\usepackage{microtype}
\usepackage{booktabs}

\ifLuaTeX
\pdfvariable pageattr{/Group <</S /Transparency /I true /CS /DeviceRGB>>}
\else
\pdfpageattr {/Group << /S /Transparency /I true /CS /DeviceRGB>>}
\fi
\title{Validating uncertainty in medical image translation}
\name{\vspace{-2.5em}} \address{
\textit{Jacob C. Reinhold$^{\star}$,
Yufan He$^{\star}$,
Shizhong Han$^{\dagger}$,
Yunqiang Chen$^{\dagger}$,
Dashan Gao$^{\dagger}$,}\\
\textit{Junghoon Lee$^{\ddagger}$,
Jerry L. Prince$^{\star \circ}$, IEEE Fellow,
Aaron Carass$^{\star \circ}$, IEEE Member}\\[1em]
$^{\star}$ Department of Electrical and Computer Engineering, Johns Hopkins
University, Baltimore,~MD,~USA \\
$^{\dagger}$ 12 Sigma Technologies, San Diego,~CA~USA \\
$^{\ddagger}$ Department of Radiation Oncology, Johns Hopkins School of Medicine, 
Baltimore,~MD,~USA \\
$^{\circ}$ Department of Computer Science, Johns Hopkins University,
Baltimore,~MD,~USA} 
\begin{document}
\maketitle
\begin{abstract}
Medical images are increasingly used as input to deep neural networks to produce 
quantitative values that aid researchers and clinicians. However, standard deep neural 
networks do not provide a reliable measure of uncertainty in those quantitative values. 
Recent work has shown that using dropout during training and testing can provide estimates of 
uncertainty. In this work, we investigate using dropout to estimate epistemic 
and aleatoric uncertainty in a CT-to-MR image translation task. We show that 
both types of uncertainty are captured, as defined, providing confidence in
the output uncertainty estimates.
\end{abstract}
\section{Introduction}
\label{sec:intro}

We build on recent developments for estimating uncertainty in deep neural 
networks~(DNNs) to give granular estimates of uncertainty in image translation 
tasks. We validate that estimates of uncertainty are captured, as defined, and 
show that uncertainty estimates can inform users about what the model knows 
and does not know, as well as provide insight into limitations inherent in 
the data and model.
 
Estimating uncertainty in DNNs is important since 
DNNs are, generally, poorly calibrated~\cite{guo2017calibration}. 
Calibration in this context refers to the average confidence in prediction diverging from model
accuracy; in the case of DNNs, the models are usually overconfident which is potentially unsafe.
Estimating uncertainty does not intrinsically calibrate a DNN, but
uncertainty\textemdash in conjunction with other measurements of model
accuracy\textemdash can improve user's trust of neural network
results.

Uncertainty has two main subtypes: \emph{epistemic} and
\emph{aleatoric}~\cite{der2009aleatory}. Epistemic uncertainty
corresponds to a model's ignorance and aleatoric uncertainty is related to the 
intrinsic variance in the data. We want to capture these two types of 
uncertainty in a medical image translation task (sometimes called image 
synthesis). Medical image translation consists of learning a function to 
transform image intensities between either two magnetic resonance~(MR) contrasts or 
two image modalities, such as $T_1$--weighted~($T_1$-w) MR and computed 
tomography~(CT)\textemdash the task we explore in this paper.

We want to estimate these two granular types of uncertainty because
1)~epistemic uncertainty shows what kind of additional training data needs 
to be acquired for optimal performance and highlights anomalies present in
the data and 2)~aleatoric uncertainty shows inherent limitations of the collected
data. Furthermore, capturing one type of uncertainty but not the other
is insufficient to estimate the \emph{predictive} uncertainty\textemdash an
encompassing measure which describes how well any voxel can be
predicted. Uncertainty estimation in image translation, segmentation, and super-resolution
has been explored~\cite{bragman2018uncertainty,nair2018exploring,tanno2019uncertainty};
however, in this work, we verify that the epistemic and aleatoric uncertainty estimates
captured in an image translation task align with the definitions of the terms.

To estimate uncertainty, we use the work of Gal and Ghahramani~\cite{gal2016dropout}, 
who show that dropout~\cite{srivastava2014dropout} can be used to learn a
variational distribution over the weights of a DNN\textemdash a form of approximate Bayesian
inference. Then, in deployment, dropout is used in a Monte Carlo
fashion to draw weights from this fitted variational distribution. The
sample variance of the output from several stochastic forward passes
corresponds to epistemic uncertainty. Aleatoric uncertainty is
captured by modifying the network architecture to create an additional
output that corresponds to a variance parameter, which is fit by
changing the loss function~\cite{kendall2017uncertainties}. We do this
by modifying a state-of-the-art supervised image translation 
DNN\textemdash a U-Net \cite{ronneberger2015unet}\textemdash to
capture the two primary types of uncertainty. 

\section{Methods}
In this section, we describe 1)~the relevant uncertainty estimation
theory and 2)~our modifications to a U-Net to estimate uncertainty.

\subsection{Uncertainty estimation}

We wish to estimate predictive uncertainty and we use the variance of the 
predicition as a proxy. Predictive uncertainty can be split into two parts which
separately estimate epistemic and aleatoric uncertainty. 

We refer the reader to Kendall and Gal~\cite{kendall2017uncertainties} for a full derivation
of the loss function and predictive variance. See Reinhold et al.~\cite{reinhold2020finding} 
for additional context to our specific method. Briefly, for paired (flattened) training data 
$\mathbf x, \mathbf y \in \mathbb R^M$, our loss function will be:
\begin{equation}
\label{eq:loss}
\mathcal{L}(\mathbf{y}, \hat{\mathbf{y}}) = \frac{1}{M} \sum_{i=1}^M
\frac{1}{2} \boldsymbol{\hat{\sigma}}_i^{-2} \norm{\mathbf{y}_i -
\mathbf{\hat{y}}_i}_2^2 + \frac{1}{2} \log \boldsymbol{\hat{\sigma}}_i^2,
\end{equation}
where $\hat{\mathbf y} = f^{\mathbf W}_{\mathbf{\hat{y}}}(\mathbf x)$
and $\boldsymbol{\hat{\sigma}}^2 = f^{\mathbf W}_{\boldsymbol{\hat{\sigma}}^2}(\mathbf x)$ 
are each outputs of a multi-task neural network $f^{\mathbf W}(\cdot)$. 
When we learn the weights, $\mathbf W$, according to Eq. (\ref{eq:loss}),
we are doing maximum likelihood estimation not only for $\hat{\mathbf{y}}$, 
but for the parameter $\hat{\boldsymbol{\sigma}}^2$, which is a per-voxel 
estimate of the data variance\textemdash a quantity related to aleatoric uncertainty.

The predictive variance of a test sample $\mathbf x^*$, with unseen target $\mathbf y^*$, can be approximated as follows:
\begin{equation*}
\mathrm{Var}(\mathbf y^*) \approx \underbrace{\frac{1}{T}
\sum_{t=1}^T \mathrm{diag}(\boldsymbol{\hat{\sigma}}^2_{(t)})}_{\text{aleatoric}}
 + \underbrace{\frac{1}{T} \sum_{t=1}^T \mathbf{\hat{y}}^2_{(t)} -
\left(\frac{1}{T}\sum_{t=1}^T \mathbf{\hat{y}}_{(t)}\right)^2}_{\text{epistemic}}\!,
\end{equation*}
where $T$ is the number of sampled weights (sampled with dropout at test time). 
Consequently, the epistemic uncertainty is the term in the predictive
variance that corresponds to sample variance while the aleatoric
uncertainty is the term associated with the mean estimated variance of the data.

\subsection{Network architecture}
We use a U-Net~\cite{ronneberger2015unet} architecture modified as follows:
\begin{itemize}
\itemsep0em 
\item We used two 3D convolutional layers, one at the start and one at the end. 
      This improved sharpness and slice-to-slice consistency.
\item We downsampled and upsampled three times instead of four. Experimental results
      showed no improvement with four downsample operations.
\item We substituted max-pooling layers for strided convolutions in
      downsampling. For upsampling we used nearest-neighbor interpolation followed by a $5^2$ convolution~\cite{odena2016deconvolution}. 
\item We attached two heads to the end of the network, where one output 
      $\hat{\mathbf{y}}$ and the other output $\hat{\boldsymbol{\sigma}}^2$. Both consisted
      of $3^3$ and $1^3$ convolutional layers.
\item We concatenated the input image to the feature maps output by the network immediately before both heads~\cite{zhao2017whole}.
\item We used spatial dropout~\cite{tompson2015efficient} ($p=0.2$)
      on all layers except the heads, 
      because it drops weights on convolutional layers
      unlike standard dropout~\cite{srivastava2014dropout}.
\item We used the AdamW optimizer \cite{loshchilov2019decoupled} with weight decay $10^{-6}$,
      learning rate 0.003, $\beta = (0.9,0.99)$, and batch size 36. 
\item We used $T=50$ weight samples in prediction.
\end{itemize}

\section{Experiments}

\begin{figure*}[!th]
\centerline{\includegraphics[width=0.28\textwidth, clip, trim=0cm 0cm
0cm 0cm]{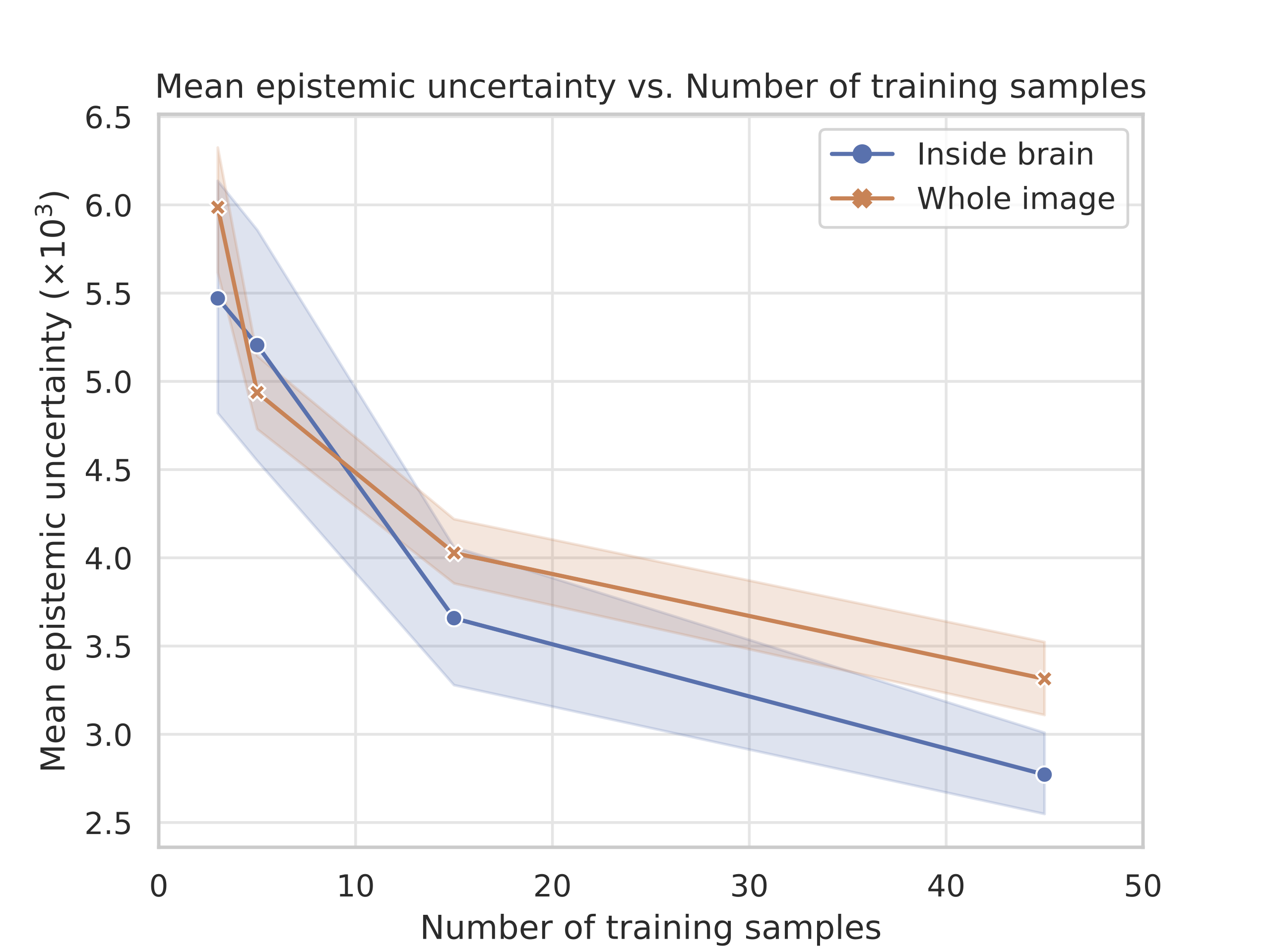} \hfil \includegraphics[width=0.28\textwidth, clip,
trim=0cm 0cm 0cm 0cm]{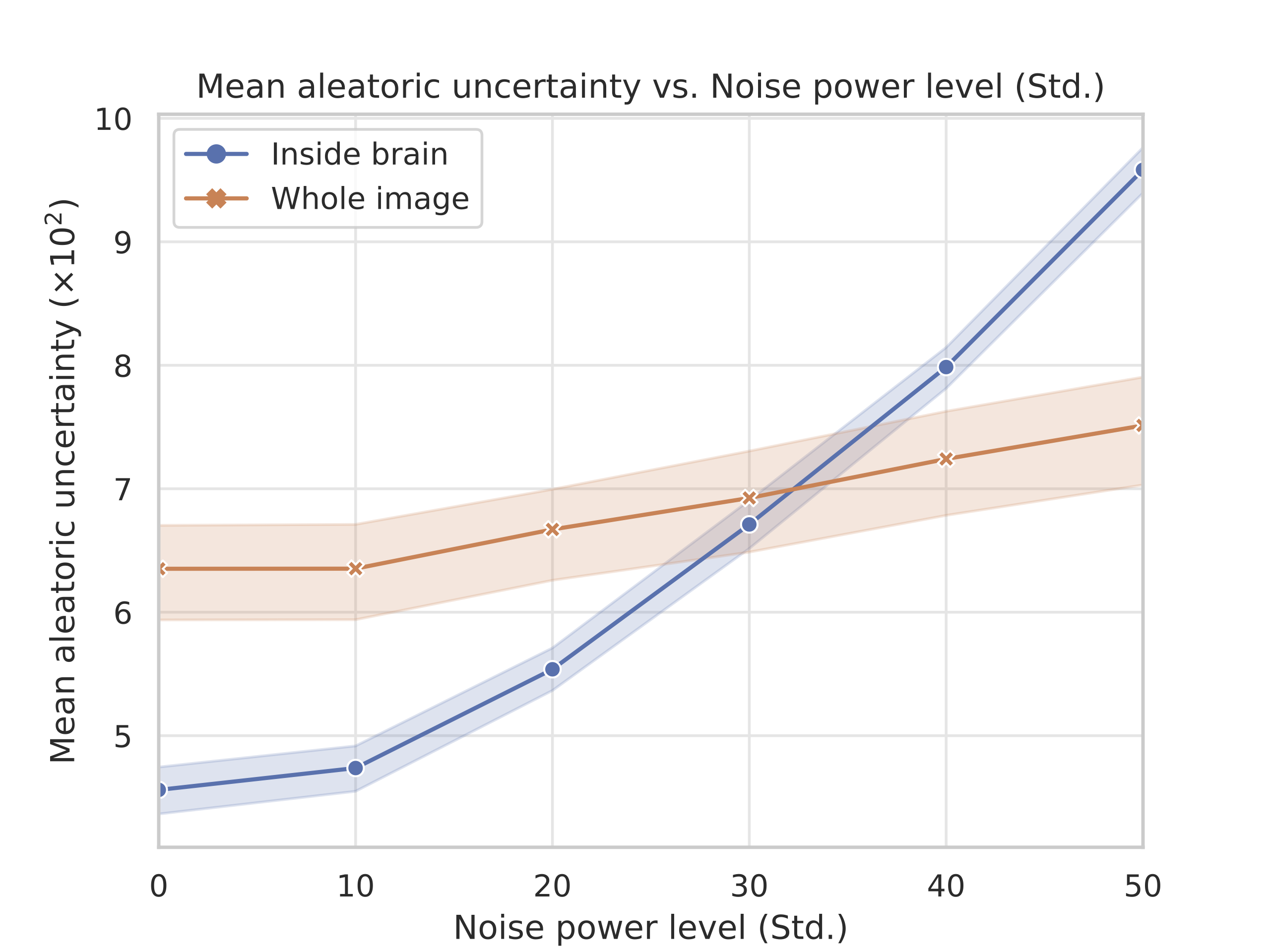}}
\caption{\label{fig:valid}\textbf{Uncertainty validation results:}
Shown are the mean epistemic~(left) and aleatoric~(right) uncertainty.
Shaded regions are bootstrapped 95\% confidence intervals.}
\end{figure*}

We conducted three experiments: 1)~an experiment to validate that we
capture both epistemic and aleatoric uncertainty, 2)~an
experiment translating CT to MR in the presence of anomalies to show that 
uncertainty is captured, as expected, in the anomalous region, and 
3)~a comparison of the synthesized $T_1$-w
image created with the proposed method and the synthesized $T_1$-w
image created with a traditional U-Net and MSE loss.

\subsection{Dataset}

We used non-contrast $T_1$-w and CT images from 51 subjects on a protocol for 
retrospective data analysis approved by the institutional review board. Fifty of
the subjects were considered to be healthy and the remaining subject
had anomalies in the brain and was excluded from training and validation. \Long{The
$T_1$-w images were acquired on a Siemens Magnetom Espree 1.5T
scanner (Siemens Medical Solutions, Erlangen, Germany, TE =
4.24 ms, TR = 1130ms, flip angle = 15$^\circ$, image size =
512$\times$512 pixels, pixel size = 0.5$\times$0.5 mm$^2$, slice
thickness = 1 mm); geometric distortions were corrected on the Siemens
Syngo application.} \Short{See Reinhold et al.~\cite{reinhold2020finding} for additional
dataset details.} All $T_1$-w images were processed to normalize the
white matter mean \cite{reinhold2019evaluating}\Long{.}\Short{ and} \Long{The CT images were acquired on a Philips Brilliance
Big Bore scanner (Philips Medical Systems, Andover, MA, image
size = 512$\times$512 pixels, pixel size = 0.6$\times$0.6 mm$^2$ –
0.8$\times$0.8 mm$^2$, slice thickness = 1.0 mm).}\Long{ All}\Short{all} images
were resampled to have a digital resolution of $0.7 \times 0.7 \times
1.0$ mm$^3$. Finally, the $T_1$-w images were rigidly registered to the CT
images. For training, the $T_1$-w and CT images were split into
overlapping $128 \times 128 \times 8$ patches. Test images were split
into three overlapping segments along the inferior-superior axis due to memory
constraints.

\subsection{Uncertainty validation on synthetic data}
\label{ss:uvsd}

To show that we capture epistemic uncertainty, we trained four networks
with 3, 5, 15, and 45 datasets. The remaining 5 healthy images are
used for validation/testing. Each network is trained until the
validation loss plateaued. In this experiment, we expect to see epistemic uncertainty
decrease on the held-out in-sample data as the training data size
increases. We see this in the left-hand plot of Fig.~\ref{fig:valid}.

To show that we capture aleatoric uncertainty, we used the network
trained on 45 datasets and added varying levels of zero-mean
Gaussian noise to the test data~(standard deviations of 10, 20, 30,
40, and 50). The right-hand plot in Fig.~\ref{fig:valid} shows the
mean aleatoric uncertainty over the entire image and the brain mask. As
expected, we see that aleatoric uncertainty increases with the level of
noise. Aleatoric uncertainty over the whole image increases more
slowly than inside the brain because, regardless of the noise level,
the network predicts relatively low aleatoric uncertainty in the background.

\subsection{Anomaly localization}

\begin{figure}[!b]
\centering
\begin{tabular}{cccc}
\textbf{CT} & \textbf{Synthesized} & \textbf{Epistemic} & \textbf{Aleatoric} \\
\includegraphics[width=0.095\textwidth, clip, trim=0.2cm 0.2cm 0.2cm 1.2cm]{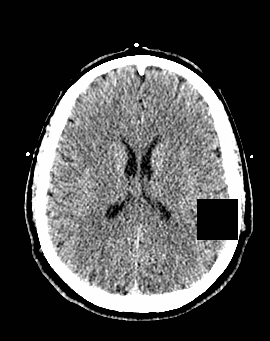} &
\includegraphics[width=0.095\textwidth, clip, trim=0.2cm 0.2cm 0.2cm 1.2cm]{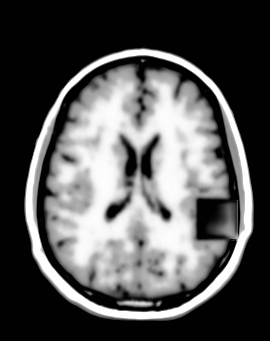} &
\includegraphics[width=0.095\textwidth, clip, trim=0.2cm 0.2cm 0.2cm 1.2cm]{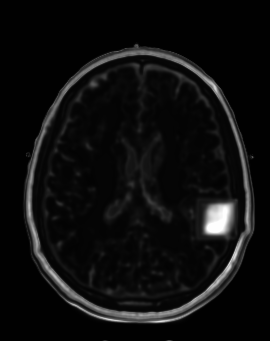} &
\includegraphics[width=0.095\textwidth, clip, trim=0.2cm 0.2cm 0.2cm 1.2cm]{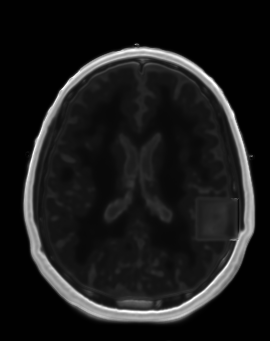}
\end{tabular}
\caption{\label{fig:epistemic}\textbf{Synthetic anomaly:}
Shown is an example synthetic anomaly CT image and the corresponding
synthetic $T_1$-w image and estimated epistemic/aleatoric uncertainty.}
\end{figure}

A DNN trained only on healthy data should exhibit high epistemic
uncertainty for input containing pathologies or, more generally,
anomalies. We quantitatively show this by inserting synthetic
anomalies into the test data and measuring the relative epistemic
uncertainty inside and outside the anomaly.

\begin{figure}[!b]
\centering
\includegraphics[width=0.47\textwidth, clip, trim=0cm 0mm 0cm 0mm]{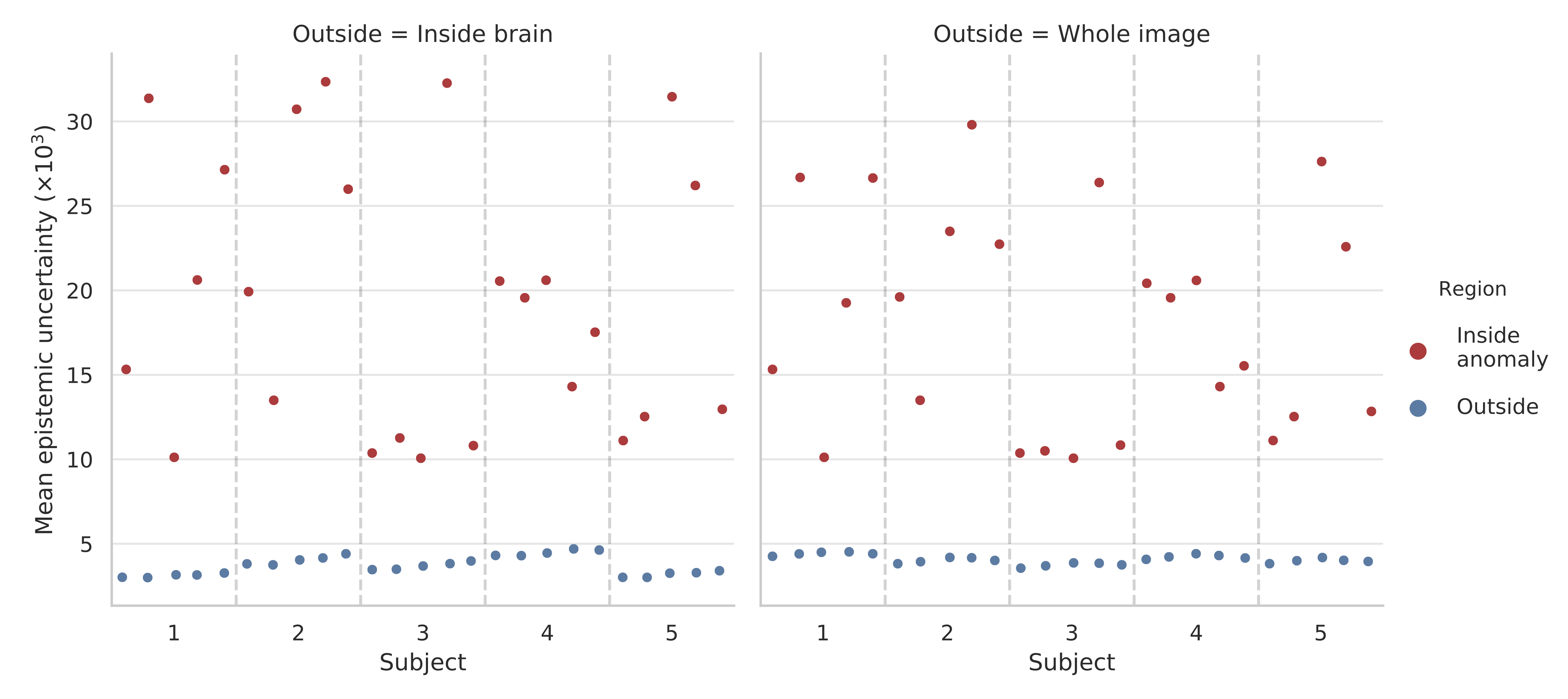}
\caption{\label{fig:quantep}\textbf{Epistemic uncertainty in synthetic anomaly}: Shown are the mean epistemic uncertainty inside
the anomaly and outside the anomaly for the 25 test images. Outside is defined
as the remainder of the brain mask~(left-hand side) or the remainder of the 
image~(right-hand side). Epistemic uncertainty inside the anomaly differs slightly 
between the two plots because the synthetic anomaly was sometimes partly outside
the brain mask.}
\end{figure}

Our synthetic anomaly is an all-zero cube of side-length 40 voxels.
This is placed randomly inside the brain mask of the five held-out
healthy CT images (see Fig.~\ref{fig:epistemic} for an example). We
create five of these anomalies per test subject by varying anomaly
location, which results in a total of 25 anomalous test images. These
synthetic anomalous data are used as input to the network trained on
15 datasets (described in Section~\ref{ss:uvsd}).
Figure~\ref{fig:quantep} shows the mean epistemic uncertainty inside
and outside the anomaly.

We also tested the model on the held-out pathological dataset collected on
the same scanner. The results are shown in Fig.~\ref{fig:comp_ex}
with the generated epistemic and aleatoric uncertainties for the input CT image.

\subsection{Comparison with U-Net}

\begin{figure}[!t]
\centering
\begin{tabular}{ccc}
\textbf{CT} & \textbf{T1-w} & \textbf{U-Net} \\
\includegraphics[width=0.11\textwidth, clip, trim=0.2cm 0cm 0.2cm 1.4cm]{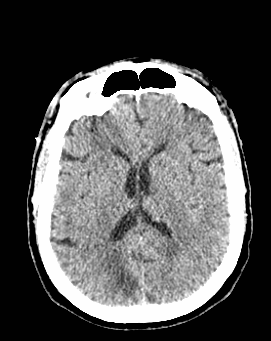} &
\includegraphics[width=0.11\textwidth, clip, trim=0.2cm 0cm 0.2cm 1.4cm]{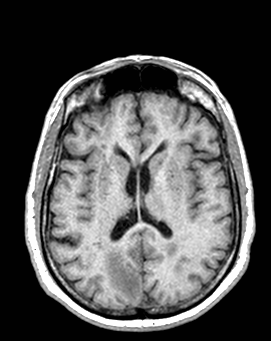} &
\includegraphics[width=0.11\textwidth, clip, trim=0.2cm 0cm 0.2cm 1.4cm]{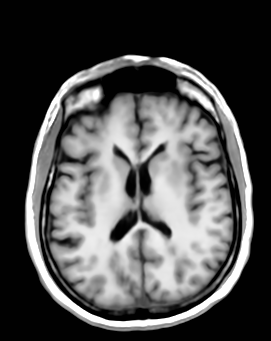} \\
\textbf{Proposed} & \textbf{Epistemic} & \textbf{Aleatoric} \\
\includegraphics[width=0.11\textwidth, clip, trim=0.2cm 0cm 0.2cm 1.4cm]{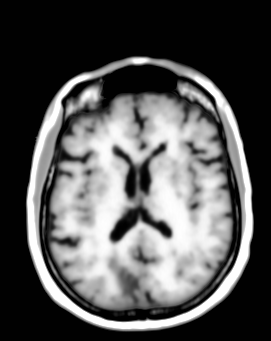} &
\includegraphics[width=0.11\textwidth, clip, trim=0.2cm 0cm 0.2cm 1.4cm]{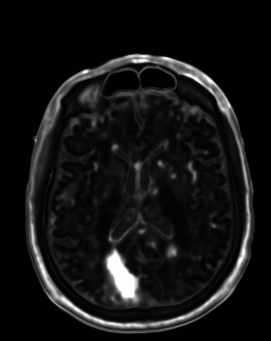} &
\includegraphics[width=0.11\textwidth, clip, trim=0.2cm 0cm 0.2cm 1.4cm]{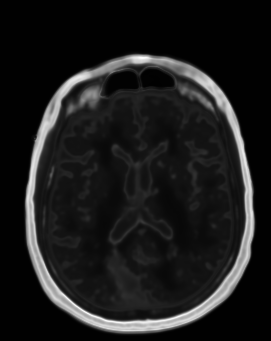} 
\end{tabular}
\caption{\label{fig:comp_ex}\textbf{Example anomalous image:}
Shown is an example anomalous image and corresponding synthesized images using
both the standard U-Net (top row, far right) and the proposed method 
(bottom row, far left). The bottom row middle and right image show the proposed 
method's additional outputs of estimated epistemic and aleatoric uncertainty maps.}
\end{figure}

In this experiment, we compare the synthesis quality of our modified
U-Net to a standard U-Net. To conduct this test, we used essentially the same network as 
previously described except 1)~there is only one head associated
with the synthesized $T_1$-w image, 2)~we did not use any dropout, and
3)~we used MSE as the loss function. We trained this U-Net and our 
proposed model on the same 45 healthy image pairs. 

Figure~\ref{fig:comp_ex} shows a qualitative result where we
examine the differences in synthesis quality on the held-out anomalous
image. Both the U-Net and proposed method fail to correctly
synthesize a portion of the brain in the occipital lobe due to the 
presence of an anomaly; however, the epistemic uncertainty map identifies that 
region as uncertain. 

Various quantitative metrics on the five held-out datasets 
show worse performance on the proposed network as compared to the U-Net. This degraded 
performance disagrees with the results by Bragman et al. \cite{bragman2018uncertainty}
who argue that the multi-task architecture regularizes training. We believe
the cause of our degraded performance is due to: 1) the high noise levels associated
with the input CT (in contrast, Bragman et al. explored MR-to-CT synthesis) and 2) that the proposed network\textemdash in spite of averaging multiple
samples\textemdash has $\sim 20\%$ fewer weights (in each pass) compared to the standard U-Net.

\section{Conclusions}

We have shown that epistemic and aleatoric uncertainty can be estimated in a medical 
image translation task with modifications to a standard DNN. As shown in prior work
(e.g., \cite{bragman2018uncertainty,tanno2019uncertainty}), the final experiment demonstrates 
how having an estimate of uncertainty adds insight into a quantitative result (see Fig.~\ref{fig:comp_ex}). 
The U-Net prediction provides no measure of uncertainty and estimates the 
structure of the anomalous brain incorrectly, which is not immediately obvious 
upon review of the synthetic image. The proposed network has a similar problem in the
synthesized image; however, because the epistemic 
uncertainty is high in the anomalous region, we know not to trust the 
corresponding synthesized values. Likewise, the uncertainty in the skin and fat
outside the skull has both high aleatoric and epistemic uncertainty
which tells us that the model has difficulty estimating the values in
those areas and that those synthesized regions cannot be trusted. 
Since we have shown that epistemic and aleatoric uncertainty is captured, as defined, 
in a medical image translation task, we can have more trust in these uncertainty estimates 
for downstream applications and analysis.

\bibliographystyle{IEEEbib}
\bibliography{biblio}

\end{document}